 \documentclass[superscriptaddress]{revtex4-1}
 \usepackage{graphicx}
 \usepackage{color}
 \usepackage[sort&compress]{natbib}
 \usepackage{mathrsfs} 
 \usepackage{amsmath,epsfig}

  \def\beq{\begin{equation}}
  \def\eeq{\end{equation}}
  \def\beqa{\begin{eqnarray}}
  \def\eeqa{\end{eqnarray}}
  \def\ban{\begin{eqnarray*}}
  \def\ean{\end{eqnarray*}}
  \def\bi{\begin{itemize}}
  \def\ei{\end{itemize}}

\begin{document}
\title{A chiral model approach to quark matter nucleation in neutron stars} 
  \author{Domenico Logoteta}
  \affiliation{Centro de F\'{\i}sica Computacional, Department of Physics,
   University of Coimbra, 3004-516 Coimbra, Portugal}
 \author{Ignazio Bombaci}
  \affiliation{Dipartimento di Fisica ``Enrico Fermi'', 
 Universit\'a di Pisa,
 and INFN Sezione di Pisa, 
  Largo Bruno Pontecorvo 3, I-56127 Pisa, Italy }
\author{Constan\c{c}a Provid\^encia}
  \affiliation{Centro de F\'{\i}sica Computacional, Department of Physics,
  University of Coimbra, 3004-516 Coimbra, Portugal}
  \author{Isaac Vida\~na}
  \affiliation{Centro de F\'{\i}sica Computacional, Department of Physics,
  University of Coimbra, 3004-516 Coimbra, Portugal}
 
 \begin{abstract}
 
 The nucleation process of quark matter in both cold and hot dense hadronic matter is investigated using a chiral approach 
 to describe the quark phase. We use the Nambu--Jona--Lasinio and the Chromo Dielectric models 
 to describe the deconfined phase and the non-linear Walecka model for the hadronic one. 
 The effect of hyperons on the transition phase between hadronic and quark matter is studied. 
 The consequences of the nucleation process for neutron star physics are outlined. 
 \end{abstract}
 
  \maketitle
  
  \vspace{0.50cm}
  PACS number(s): {97.60.s, 97.60.Jd, 26.60.Dd, 26.60.Kp} 
  \vspace{0.50cm}


  \section{Introduction}
 
In the last few years the physics of dense matter is having a growing interest due to its applications in both terrestrial (heavy ion collisions)  and astrophysical (supernovae explosions and neutron stars) systems. In particular, the question of the existence of quark matter in the universe is 
still an open problem. Neutron stars are among the densest objects in the universe, and,  therefore, they are probably one of the best candidates to host a quark deconfined phase in their cores. 
 
Quark matter nucleation in neutron star has already been discussed in several previous works.
Zero temperature nucleation was studied in Refs.\ \cite{b0,b1,b2,b3,b4,b5,b6,b7,b8}, while in 
Refs.\ \cite{h1,h2,h3,h4,me,me1} thermal nucleation was considered. 
In all these papers the MIT bag model \cite{mit} has been used for the description of quark phase. Recently, the nucleation rate and 
the nucleation time have been calculated at zero and finite temperature within the same quark model \cite{me,me1}. This calculation has allowed to follow the thermal evolution of neutron stars  
from the young and warm proto-neutron stars to the cold ($T = 0$) and deleptonized neutron stars.
The crossover temperature above which the thermal nucleation dominates quantum nucleation was
calculated and the consequences of the possible quark nucleation for the physics and evolution of
proto-neutron stars were discussed. It was shown that proto-hadronic stars with a gravitational
mass below a critical value will survive the early stages of their evolution without 
decaying into a quark star.
The effect of neutrino trapping on quark nucleation has been considered 
explicitly in Refs.\ \cite{lug1} and \cite{me1}.
 
In the present paper we will investigate how the nucleation process depends on the quark model
chosen to describe the dense quark phase. There are several chiral quark models in the literature 
that successfully  describe the low energy properties of mesons and hadrons. In particular, we will consider two quark models which contain explicitly the chiral symmetry: 
the Nambu--Jona--Lasinio (NJL) model \cite{nambu} and the Chromo Dielectric Model (CDM) \cite{cdm}.  
The NJL model (see \cite{bba2} for a recent review) has been extensively used to study quark 
matter in $\beta$-equilibrium and quark stars \cite{bbub,cost,bba,bba1,bba2,benh1}.  
The parameters of this model 
are fixed by low energy scattering data of mesons.  
Although, the NJL model contains dynamical chiral symmetry breaking explicitly, an important  
symmetry of QCD, it is unable to explain confinement.
This feature is maybe the main drawback of the model. 
 
The QCD phase diagram determined with the NJL model, shows that, at very high densities and low temperatures, the most favourable  deconfined phase could be in a color superconducting state 
in which two flavors or all three quark flavors are paired \cite{bla,rust}. 
High densities and low temperatures are the conditions which are expected to occur at the core of a neutron star some minutes after its birth. In the present  paper we do not take into account the 
possible formation of a color-superconducting state of dense quark matter and  
we use the standard version of the NJL model without superconductivity \cite{njll}.  
Recently the deconfinement phase transition in proto-neutron stars within  
the SU(3) NJL  model including 2SC colour superconductivity has been discussed by the authors of 
Refs. \cite{lug} and \cite{lug1}.   

 The CDM has been used to study the static and dynamical properties of the nucleon, and describes the confinement of quarks through their
 interaction with a scalar field $\chi$ which represents a multiâ€“gluon state and produces a density
 dependent constituent mass. Quark matter has been analysed within this model in several papers \cite{cdm2,cdmqm,cdm7}.
  
 The present work is organized as follows:
 first we give a brief description of the approach used to model the hadronic phase; next
 we review the main features of the NJL and Chromo Dielectric models and present the results 
 for quark matter nucleation in both cold and hot dense hadronic matter. Finally, the 
 consequences of the deconfinement phase transition for neutron star physics are discussed. 
 
 
 \section{Equation of state}
 
 In this section we present a brief review of the models used to build the equation of
 state (EOS) of stellar matter. The hadronic matter is described within the non-linear
 Walecka model (NLWM) \cite{nlwm} and quark matter within the NJL and the CDM models. We take
 $\hbar=c=1$ in all the expressions. 
 
 
 \subsection{The non-linear Walecka model}
 
   The Lagrangian density, including the baryonic octet, in terms of the scalar
   $\sigma$, the vector-isoscalar $\omega_\mu$ and the vector-isovector $\vec \rho_\mu$
   meson fields reads (see {\it e.g.} \cite{prak97,glen00,cost})  
  \begin{equation}
  {\cal L}={\cal L}_{hadrons}+{\cal L}_{leptons}
  \label{lagB}
  \end{equation}
  where the hadronic contribution is
  \begin{equation}
  {\cal L}_{hadrons}={\cal L}_{baryons}+{\cal L}_{mesons}
  \end{equation}
  with
  \begin{equation}
  {\cal L}_{baryons}=\sum_{B} \bar \psi_B\left[\gamma^\mu D_\mu -M^*_B\right]\psi_B,
  \end{equation}
  where
  \begin{equation}
  D_\mu=i\partial_{\mu}
  -g_{\omega B} \omega_{\mu}-{g_{\rho B}} \vec{t}_B \cdot \vec{\rho}_\mu, 
  \end{equation}
  and 
  $M^*_B=M_B-g_{\sigma B} \sigma$ is the baryon effective mass. The quantity $\vec{t}_B$ designates the isospin of
  baryon $B$.
  The mesonic contribution reads 
  \begin{equation}
  {\cal L}_{mesons}={\cal L}_{\sigma}+{\cal L}_{\omega}+ {\cal L}_{\rho},
  \end{equation}
  with
  \begin{equation}
      {\cal L}_\sigma=\frac{1}{2}(\partial_{\mu}\sigma\partial^{\mu}\sigma
      -m_{\sigma}^2 \sigma^2)+ \frac{1}{3!} \kappa \sigma^3+ \frac{1}{4!} \lambda
      \sigma^4,
  \end{equation}
  \begin{equation}
      {\cal L}_{\omega}=-\frac{1}{4}\Omega_{\mu\nu}\Omega^{\mu\nu}+\frac{1}{2}
      m_{\omega}^2 \omega_{\mu}\omega^{\mu}, \qquad \Omega_{\mu\nu}=\partial_{\mu}\omega_{\nu}-\partial_{\nu}\omega_{\mu},
  \end{equation}
  \begin{equation}
      {\cal L}_{\rho}=
     { -\frac{1}{4}\vec B_{\mu\nu}\cdot\vec B^{\mu\nu}}+\frac{1}{2}
      m_\rho^2 \vec \rho_{\mu}\cdot \vec \rho^{\mu}, \quad \vec B_{\mu\nu}=\partial_{\mu}\vec \rho_{\nu}-\partial_{\nu} \vec \rho_{\mu}
        - g_\rho (\vec \rho_\mu \times \vec \rho_\nu) \;.
  \end{equation}
  For the lepton contribution we take
  \begin{equation}
  {\cal L}_{leptons}=\sum_{l} \bar \psi_l \left(i \gamma_\mu \partial^{\mu}-
  m_l\right)\psi_l,
  \end{equation}
  where the sum is over electrons, muons 
 and neutrinos for matter with trapped neutrinos. 
  In uniform matter, we get for the baryon Fermi energy
  $
        \epsilon_{FB}=g_{\omega B} \omega_0+ g_{\rho B} t_{3B} \rho_{03} + \sqrt{k_{FB}^2+{M^*_B}^2}.
  $
  
 We will use the GM1 parametrization of NLWM due to Glendenning and Moszkowski \cite{gm91,glen00}. The nucleon coupling constants are fitted to the bulk properties of nuclear matter. 
The inclusion of hyperons involves  new couplings, which can be written in terms of the nucleonic ones: $g_{\sigma Y}=x_{\sigma }~ g_{\sigma},~~g_{\omega Y}
 =x_{\omega }~ g_{\omega},~~g_{\rho Y}=x_{\rho }~ g_{\rho}$. In this model it is assumed that all the hyperons in the  octet have the same coupling. 
 Measured neutron star masses can be used to restrict the possible ranges 
of variability of the hyperon couplings \cite{gm91,glen00}. In this work we will consider 
$x_\sigma=0.6$ and $x_\sigma=0.7$.   
In addition, following Ref. \cite{gm91} we will take  $x_{\rho} = x_{\sigma}$, whereas 
the binding energy of the $\Lambda$ in symmetric nuclear matter, $B_\Lambda$,
\begin{equation}
 \left(\frac{B_\Lambda}{A}\right)=-28 \mbox{ MeV}= x_{\omega} \, g_{\omega}\, \omega_0-x_{\sigma}\,  g_{\sigma} \sigma 
\end{equation}
is used to determine $x_\omega$ in terms of $x_\sigma$.  
Notice that the case with $x_{\sigma} = 0.6$ produces stars with a larger hyperon population (for a given stellar gravitational mass) with respect to the case  $x_{\sigma} = 0.7$  
\cite{glen00,b6}. In addition to these two parametrizations for hyperonic matter, 
we will consider the case of pure nucleonic matter (hereafter called $np$ matter).  

At zero temperature the equations of motion for the various fields can be derived using the Lagrangian of Eq. (\ref{lagB}) and 
 the well known Euler-Lagrange equations. At finite temperature we use the 
following grand canonical potential per unit volume:
\begin{eqnarray} 
\Omega &=& \frac{1}{2} m_{\sigma}^2 \sigma^2 
+ \frac{1}{3!} k \sigma^3 + \frac{1}{4!} \lambda \sigma^4 
- \frac{1}{2} m_{\omega}^2 \omega_0^2                      \nonumber \\
\nonumber \\
&-& \frac{1}{2} m_{\rho}^2 \rho_{03}^2 -  2T \sum_{i=B} \int \frac{d^3 k}{(2\pi)^3} \left\{  
\ln \left[ 1 + e^{-\beta(E_i^* - \nu_i)} \right]  \right.  \nonumber \\
\nonumber \\
&+& \left. \ln \left[ 1 + e^{-\beta(E_i^* + \nu_i)}  \right]  \right\}  \;,  
\end{eqnarray}
where $\beta = 1/T$, $E_i^* = ({\mathbf k}_i^2 + M_i^{*\,2})^{1/2}$ 
and the effective chemical potential of baryon $i$ is given by
\begin{equation}
  \nu_i = \mu_i - g_{\omega}\omega_0 - \tau_{3i}  \, g_{\rho}\, \rho_{03} \;.
\end{equation}
The sum is extended to all baryons of the baryonic octect.
The equations of motion are then obtained minimizing $\Omega$ with respect to the fields $\sigma$, $\omega_0$ and $\rho_{03}$. The 
baryonic pressure and energy density are given by:
\begin{equation}
P_B=-\Omega \;, \ \ \ \ \ \ \varepsilon=-P_B+\sum_{i=B} \mu_i n_i+s_B T
\end{equation}    
where $n_i$ is the particle number density of the baryon $i$ and $s_B$ is the total baryonic entropy density:
\begin{equation}
s_B=-\left(\frac{\partial \Omega}{\partial T}\right)_{\mu_i, V} \;.
\end{equation} 
The total energy density and pressure are finally obtained adding the lepton contribution to $P_B$, $\varepsilon_B$ and $s_B$. 

 
 \subsection{The Nambu--Jona--Lasinio model}
 
 The three flavor NJL model \cite{nambu} has been widely used by many authors to describe 
quark matter \cite{bba2,bbub,cost,bba,bba1,benh1}. In this paper we adopt the Lagrangian 
density of Ref. \cite{njll}:
 
 \begin{equation}
 {\cal L}=\bar{q}(i \gamma_\mu\partial^\mu-m_{0})q+G\sum_{k=0}^{8}[(\bar{q}\lambda_{k}q)^{2}+(\bar{q}i\gamma_{5}\lambda_{k}q)^{2}]-
 K[det_{f}(\bar{q}(1+\gamma_{5})q)+det_{f}(\bar{q}(1-\gamma_{5})q)] \ .
 \end{equation}
 \label{lagnjl}
 In the above expression, $q$ denotes the three flavor quark field: $q=(u,d,s)$ and $m_{0}=diag(m_{0}^{u},m_{0}^{d},m_{0}^{s})$ is the mass matrix.
 The model is not renormalizable and we use a sharp cut off $\Lambda$ to treat the divergent integrals. Following \cite{njll} we take: $\Lambda=602.3$ MeV,
  $G \Lambda^{2}=1.835$, $K \Lambda^{2}=12.36$, $m_{0}^{u}=5.5$ MeV, $m_{0}^{d}=5.5$ MeV, $m_{0}^{s}=140.7$ MeV. These parameters have been 
 determined by fitting the $f_{\pi}$ (pion decay constant), $m_{\pi}$ (pion mass), $m_{K}$ (kaon mass) and $m_{\eta^{'}}$ ($\eta'$ mass) to their experimental 
 values. Due to the interaction,
  the mass of the quarks in the medium (dynamical masses) are in general different from the bare masses (current masses). In the Hartree approximation, 
 the dynamical quark masses are determined by the solution of the gap equation:
 \begin{equation}
 \label{g1}
 m_{i}=m_{0}^{i}-4G<\bar{q_{i}}q_{i}>+2K<\bar{q_{j}}q_{j}><\bar{q_{k}}q_{k}>, \,\,\, i \neq j \neq k \, .
 \end{equation}
 The quark condensates $<\bar{q_{i}}q_{i}>$ at zero temperature are given by:
 \begin{equation}
 \label{g2}
 <\bar{q_{i}}q_{i}>=-\frac{3}{\pi^{2}}\int_{k_{F_{i}}}^{\Lambda} dk k^{2} \frac{m_{i}}{\sqrt{m_{i}^{2}+k^{2}}},
 \end{equation}
 where $k_{F_{i}}=(\pi^2 n_i)^{1/3}$ is the Fermi momentum of the quark $i$ and
 $n_i$ its number density. The energy density ($\varepsilon$)
 and pressure ($P$) of the system are given by 
 \begin{equation}
 \label{e0}
 \varepsilon=\sum_{i=u,d,s}\frac{3}{\pi^{2}}\int_{0}^{k_{F_{i}}}dk k^{2}\sqrt{m_{i}^{2}+k^{2}}+ B_{eff},
 \end{equation}
 \begin{equation}
 \label{e2}
 P=-\varepsilon + \sum_{i=u,d,s} n_{i}\sqrt{m_{i}^{2}+k_{F_{i}}^{2}}.
 \end{equation}
 In the NJL model the bag constant is not a phenomenological input parameter, like in the MIT bag model. However, 
 one can still define an effective bag pressure, $B_{eff}=B_0-B$, generated dynamically with origin in the spontaneous 
 breaking of chiral symmetry, where 
 \begin{equation}
 B= \sum_{i=u,d,s}\left[\frac{3}{\pi^{2}}\int_{0}^{\Lambda}dk k^{2} \sqrt{m_{i}^{2}+k^{2}}
 -2G<\bar{q_{i}}q_{i}>^{2} \right]+4K<\bar{u}u><\bar{d}d><\bar{s}s> \ ,
 \end{equation}
 and $B_{0}=B(n_{u}=n_{d}=n_{s}=0)$ is introduced to ensure zero pressure at zero density and temperature. 
 
 The system of equations (\ref{g1}), (\ref{g2}) is solved numerically for a fixed value of the
 baryonic density $\rho_{B}=(n_{u}+n_{d}+n_{s})/3$ and the pressure and energy density are then
 calculated from equations (\ref{e0}) and (\ref{e2}). 


 At finite temperature $T$, the above expressions can be generalized starting from the grand canonical potential per unit volume $\Omega$ \cite{bba2}:
\begin{equation}
\Omega=\sum_{i=u,d,s} \Omega_{m_i}+2 G <\bar{q_{i}}q_{i}>^{2}- 4K<\bar{u}u><\bar{d}d><\bar{s}s>+B_0 \; ,
\end{equation}
where 
\begin{equation}
\Omega_{m_i}= -  \frac{3}{\pi^2} \int dk \ k^2 \left[ \sqrt{k^2+m_i^2} +T \  \ln \left( 1 + e^{ - \left( \sqrt{ k^2 + m_{i}^{2}}-\mu_i \right)/T}\right)+
T \ \ln\left( 1 + e^{ - \left(\sqrt{k^2+m_{i}^{2}}+\mu_i\right)/T}\right) \right] \;.
\end{equation} 
The particle density of the quark $i$ is given by: 
\begin{equation}
n_i=\frac{3}{\pi^2} \int dk \ k^2 [f_i(k)-\bar{f}_i(k)]  \;,
\label{densities}
\end{equation}
while quark condensate becomes:
 \begin{equation}
  <\bar{q_{i}}q_{i}>=\frac{3}{\pi^{2}}\int dk \ k^{2} \frac{m_{i}}{\sqrt{m_{i}^{2}+k^{2}}} [f_i(k)+\bar{f}_i(k)-1],
\label{g2T}
\end{equation} 
being $f_i(k)$ and $\bar{f}_i(k)$ the quark and antiquark Fermi distribution functions, respectively:
\begin{eqnarray}
f_i(k)=\left(1 + e^{ \left( \sqrt{k^2+m_{i}^{2}} - \mu_i \right)/T}\right)^{-1} \\
\bar{f}_i(k)=\left(1 + e^{\left(\sqrt{k^2+m_{i}^{2}}+\mu_i\right)/T}\right)^{-1} \;.
\label{distrfunc}
\end{eqnarray} 
The gap equations are derived minimizing $\Omega$ with respect to the constituent quark masses $m_i$. The expressions obtained are identical to 
Eqs. (\ref{g1}) with the quark condensate given by Eq. (\ref{g2T}).
The pressure and entropy density of the system are given by:
\begin{equation}
P=-\Omega \;, \ \ \ \ \ \ s=-\left(\frac{\partial \Omega}{\partial T}\right)_{\mu_i, V} \;.
\label{pressure}
\end{equation}
Finally the the energy density reads:
\begin{equation}
\varepsilon=-P+\sum_{i=u,d,s} n_i \ \mu_i+ s T  \; .
\label{energy}
\end{equation}

 
 \subsection{The Chromo Dielectric Model}
 
 The Chromo Dielectric model \cite{cdm} is a confinement model that has been extensively used to study properties of single nucleons or to investigate 
 quark matter in neutron stars \cite{cdm1,cdm2,cdm3,cdmf,cdm4,cdmbaldo,cdm5,cdm6} and supernovae
 explosions \cite{cdm7,cdm8,cdm9}. Confinement is achieved through the introduction of a scalar-isoscalar chiral singlet field $\chi$.
 The Lagrangian density of the model reads: 
 \begin{equation}
 {\cal L}=i \sum_{f=u,d,s}\bar{\psi}_{f}\gamma^{\mu}\partial_{\mu}\psi_{f}+\frac{1}{2}(\partial_{\mu}\sigma)^{2}
 +\frac{1}{2}(\partial_{\mu}\vec\pi)^{2}-U(\sigma,\vec\pi)+ 
 \sum_{f=u,d}\frac{g_{f}}{f_{\pi}\chi}\bar{\psi}_{f}(\sigma+i\gamma_{5}\vec\tau \cdot \vec\pi)\psi_{f}
 +\frac{g_{s}}{\chi}\bar{\psi}_{s}\psi_{s}+
 \frac{1}{2}(\partial_{\mu}\chi)^{2}-V(\chi) \;,
 \end{equation}
 where $\psi_f$ represents the quark field of flavor $f$, $U(\sigma,\vec{\pi})$
 is a mexican-hat potential
 \begin{equation}
 U(\sigma,\vec{\pi})=\frac{m_{\sigma}^{2}}{8 f_{\pi}^{2}}(\sigma^2+\pi^{2}-f_{\pi}^{2}) \ ,
 \end{equation}
 and, for $\chi$ we consider the simplest potential
 \begin{equation}
 V(\chi)=\frac{1}{2} M_\chi^{2} \chi^{2} \;.
 \end{equation}

 The characteristic feature of the CDM is that quark masses rescale as an inverse power of the
 field $\chi$ and, therefore, acquire a density dependence
 \begin{equation}
 m_{u,d}=\frac{-g_{u,d} \sigma}{\chi f_{\pi}} \ , \ \ \ \ \ \ \ \
 m_{s}=\frac{g_{s}}{\chi} \ .
 \label{ccdm1}
 \end{equation} 
 In vacuum $\chi$ vanishes thus providing confinement.
 The coupling constants are given by $g_{u,d}=g(f_{\pi}\pm\xi_{3})$ and $g_{s}=g(2f_{K}-f_{\pi})$, where $f_{\pi}=93$ MeV and $f_{K}=113$ MeV
  are the pion and the kaon decay constants. The other two constants $\xi_{3}$ and $m_{\sigma}$ are fixed in such a way that: $\xi_{3}=f_{K_{\pm}}-f_{K^{0}}=-0.75$ MeV
  and $m_{\sigma}=1.2$ GeV. Following Ref.\ \cite{cdm7}, for the mass $M_\chi$ of the field $\chi$ and its coupling $g$ we take $M_\chi=1.7$ GeV and
 $g=23$ MeV, respectively. These values lead to reasonable values for the average delta-nucleon mass and for the nucleon isoscalar radius \cite{cdm7}. 
 
 At zero temperature the energy density in the CDM is given by:
 %
\begin{equation}
 \varepsilon=\frac{3}{\pi^2} \sum_{i=u,d,s} \int_{0}^{k_{F_i}} dk \ k^2\sqrt{k^{2}+m_{i}^{2}(\sigma, \chi)}+B_{eff},
\end{equation}
 with  
\begin{equation}
        B_{eff}=V(\chi)+U(\sigma, \vec \pi=0) \ .
\end{equation}
 In mean field approximation, the field $\vec \pi$ vanishes while the other two fields $\sigma$ and $\chi$ are replaced by their mean values. 
 The equations of motion are obtained by minimizing the energy density of the system. One gets:
 \begin{equation}
 \frac{\partial V(\chi)}{\partial \chi}=-\sum_{i=u,d} \rho_{S_i}(k_{F_i},m_{i})\frac{\sigma g_{i}}{f_{\pi}\chi^{2}}+
 \rho_{S_s}(k_{F_s},m_{s})\frac{g_{s}}{\chi^{2}} \ ,
 \end{equation}
 \label{ga1}
 \begin{equation}
 \frac{\partial U(\sigma,\vec \pi=0)}{\partial \sigma}=\sum_{i=u,d} \rho_{S_i}(k_{F_i},m_{i})\frac{g_{i}}{f_{\pi}\chi} \ ,
 \end{equation}
 \label{ga2}
 with
 \begin{equation}
 \rho_{S_i}(k_{F_i},m_{i})=\frac{3}{\pi^2} \int_{0}^{k_{F_i}} dk \ k^2 \frac{m_{i}}{\sqrt{m_{i}^{2}+k^{2}}} .
 \end{equation}
 \label{ga3}
The inclusion of temperature in the CDM has been carried out following Ref.\ \cite{cdm2}.
We can start from the grand canonical potential per unit volume $\Omega$ : 
\begin{equation}
\Omega=\sum_{i=u,d,s} - T \ \frac{3}{\pi^2} \int dk \ k^2 \  \left[ \ln \left( 1 + e^{ - \left( \sqrt{ k^2 + m_{i}^{2}}-\mu_i \right)/T}\right)+
\ln\left( 1 + e^{- \left(\sqrt{k^2+m_{i}^{2}}+\mu_i\right)/T}\right) \right] + B_{eff} \;,
\end{equation} 
where $m_i$ and $\mu_i$ are the effective mass (defined in Eq. (\ref{ccdm1})) and the chemical potential
 of the quark $i$. 
Minimizing $\Omega$ with respect to $\sigma$ and $\chi$, we get the correspondig equations of motion at $T \neq 0$.
\begin{equation}
\left(\frac{\partial \Omega}{\partial \sigma}\right)_{\mu_i, T}=0 \;, \ \ \ \ \ \ \ \left(\frac{\partial \Omega}{\partial \chi}\right)_{\mu_i, T}=0 \;.
\end{equation}
The particle densities of the three quarks are given by Eq. (\ref{densities}) and   
the pressure and energy density can be calculated using Eqs. (\ref{pressure}) and (\ref{energy}). 
 
 \section{Phase equilibrium}

In the region of high density (high baryon chemical potential) 
and low temperature (which is the one relevant for neutron star physics) many 
QCD-inspired models suggest the  deconfinement transition to be a first-order 
phase transition \cite{hs98,fk04}). 
Under this assumpion the conditions for phase equilibrium are thus  given 
by the Gibbs phase rule 
\begin{equation}
 T_H = T_Q \equiv T \, ,~~~~~~~~
 P_H = P_Q \equiv P_0 
 \label{eq:eq1a}
 \end{equation}
 \begin{equation}
 \mu_H(T, P_0) = \mu_Q(T, P_0) \, 
 \label{eq:eq1b}
 \end{equation}
 where 
 \begin{equation}
   \mu_H = \frac{\varepsilon_H + P_H - s_H T}{n_H} \, , ~
   \mu_Q = \frac{\varepsilon_Q + P_Q - s_Q T}{n_Q} 
 \label{eq:eq2}
 \end {equation} 
 are the Gibbs energies per baryon (average chemical potentials) for the hadron (H) and 
 quark (Q) phase respectively, 
 $\varepsilon_H$ ($\varepsilon_Q$), $P_H$ ($P_Q$), $s_H$ ($s_Q$) and $n_{H}$ ($ n_{Q}$)
 denote respectively the total ({\it i.e.,} including leptonic contributions) energy 
 density, total pressure, total entropy density, and baryon number density of the two phases. 
 The pressure $P_0$ defines the transition point. For pressures above $P_0$ the hadronic phase is 
 metastable, and the stable quark phase will appear as a result of a nucleation process. 
 
 Small localized fluctuations in the state variables of the metastable hadronic phase 
 will give rise to virtual drops of the stable quark phase. These fluctuation are characterized 
 by a time scale $\nu_0^{-1} \sim 10^{-23}$ s set by the strong 
 interaction that is responsible for the deconfinement phase transition. This time scale
 is many orders of magnitude smaller than the typical time scale set by weak interactions, 
 therefore, quark flavor must be conserved during the deconfinement transition. 
 We will refer to this form of deconfined quark matter, in which the flavor content is equal to that of 
 the $\beta$-stable hadronic system at the same pressure and temperature, as the $Q^*$ phase. 
 Soon afterward a critical size drop of quark matter is formed, the weak interactions 
 will have enough time to act, changing the quark flavor fraction of the deconfined droplet to lower 
 its energy, and a droplet of $\beta$-stable quark matter is formed (hereafter the $Q$ phase).
 
 This first seed of quark matter will trigger the conversion \cite{oli87,hbp91,grb} of the pure hadronic star to a quark star (hybrid star or strange star). Thus, pure hadronic stars with 
values of the central pressure larger than  $P_0$ are metastable to the decay (conversion) 
to hybrid stars or to strange stars \cite{b0,b1,b2,b3,b4,b5,b6,b7}. 
The mean lifetime of the metastable stellar configuration is related to the time needed to 
nucleate the first drop of  quark matter in the stellar center and depends dramatically 
on the value of the  stellar central pressure \cite{b0,b1,b2,b3,b4,b5,b6,b7} 
and central temperature \cite{me,me1}.

In order to explore the astrophysical implications of quark matter nucleation,  following 
Ref. \cite{b0,b1,b2}, we introduce the concept of {\it critical mass} $M_{cr}$ for the 
hadronic star sequence.

In the case of cold and deleptonized stars, $M_{cr}$ can be defined as the value  
of the gravitational mass of the metastable hadronic star for which the nucleation time $\tau$ 
takes a ``reasonable small'' value in comparison with typical ages of young pulsars as, 
{\it e.g.} the Crab pulsar.  Thus, according to Ref. \cite{b0,b1,b2} we take 
$M_{cr}(T=0) \equiv M(\tau=1~{\rm yr}, \,T=0)$.      
It is worth recalling that the nucleation time is an extremely steep function of the   
hadronic star mass \cite{b0,b1,b2}, therefore the exact value of $\tau$ chosen in the definition 
of $M_{cr}(T=0)$ is not crucial. We have verified that changing  $\tau$ from 1~yr  to $10^3$~s modifies $M_{cr}(T=0)$ by $\sim 0.02\%$.

In the case of newly formed compact stars, the characteristic evolutionary time-scale is
the proto-hadronic star cooling time $t_{cool}$, {\it i.e.} the time it takes the new born star to reach a cold and deleptonized configuration. The cooling time has been evaluated 
to be \cite{BurLat86} $t_{cool} \sim$~a~few~$10^2$~s.  
Thus, according to Ref. \cite{me,me1}, we consider isoentropic stellar configurations 
(with an entropy per baryon $\tilde{S}$ in the range 1 -- 2 $k_B$) and define the critical mass 
for proto-hadronic stars as $M_{cr}(\tilde{S}) \equiv M(\tau = 10^3~{\rm s},~\tilde{S}=$~const).

Notice that  pure hadronic stars with $M_{HS} > M_{cr}$ are very unlikely to be observed, 
while pure hadronic stars with $M_{HS} < M_{cr}$ are safe with respect to a sudden transition 
to quark matter. Thus $M_{cr}$ plays the role of an {\it effective maximum mass for the hadronic branch of  compact stars} (see discussion in Ref.\ \cite{b2}). 
While the Oppenheimer--Volkov maximum mass is determined by the overall stiffness of the equation of state for hadronic matter, the value of $M_{cr}$ will depend in addition on the properties of the intermediate non $\beta$-stable $Q^*$ phase.

 
\section{Quark matter nucleation rates}
 
The nucleation process of quark matter in hadronic stars can take place during  different stages of their evolution \cite{me1}.   
This is due to the fact that nucleation can proceed both via thermal activation or quantum 
tunnelling (at zero or finite temperature).    
The core of a newborn neutron star reaches temperatures of $10 - 40$ MeV 
\cite{BurLat86,prak97} and, consequently, this era is dominated by the thermal nucleation 
regime;  on the other hand a cold deleptonized neutron star can nucleate quark matter 
only via quantum tunnelling because the thermal nucleation time diverges in the limit of zero temperature (see \cite{me,me1} and the following discussion).

The energy barrier, which represents the difference in the free energy of the system with and 
 without a $Q^*$-matter droplet, can be written as 
 \begin{equation}
   U({\cal R}, T) = \frac{4}{3}\pi n_{Q^*}(\mu_{Q^*} - \mu_H){\cal R}^3 + 4\pi \sigma {\cal R}^2
 \label{eq:potential}
 \end{equation}
 where ${\cal R}$ is the radius of the droplet (supposed to be spherical), and $\sigma$ is 
 the surface tension for the surface separating the hadronic phase from the $Q^*$ phase. 
 The energy barrier has a maximum at the critical radius 
 ${\cal R}_c = 2 \sigma /[n_{Q^*}(\mu_H - \mu_{Q^*})]$. 
 Notice that we have neglected the term associated with the curvature energy, 
 and also the terms connected with the electrostatic energy, since they are known to introduce only small corrections \cite{b2,iida98}. 
 The value of the surface tension $\sigma$ for the interface separating the quark and hadronic phase 
 is poorly known, and the values typically used in the literature range within $10-50$ MeV fm$^{-2}$ 
 \cite{hei93,iida97,iida98}. In the following, we assume $\sigma$ to be temperature independent and we take 
 $\sigma = 30$ MeV fm$^{-2}$. 
 
 The quantum nucleation time $\tau_q$ (at zero and finite temperature) can be straightforwardly evaluated within a semi-classical 
 approach \cite{lk72,iida97,iida98}. First one computes, in the Wentzel--Kramers--Brillouin (WKB) approximation, the ground state 
 energy $E_0$ and the oscillation frequency $\nu_0$ of the drop in the potential well $U({\cal R},T)$. 
 Then, the probability of tunnelling is given by
 \begin{equation}
   p_0=exp\left[-\frac{A(E_0)}{\hbar}\right]
 \label{eq:prob}
 \end{equation}
 where $A(E)$ is the action under the potential barrier, which in a relativistic framework reads
 \begin{equation}
  A(E)=\frac{2}{c}\int_{{\cal R}_-}^{{\cal R}_+}\sqrt{[2m({\cal R})c^2 +E-U({\cal R})][U({\cal R})-E]} 
 \label{eq:action}
 \end{equation} 
 being ${\cal R}_\pm$ the classical turning points and $m({\cal R})$ the droplet effective mass. 
 The quantum nucleation time is then equal to
 \begin{equation}
   \tau_q = (\nu_0 p_0 N_c)^{-1} \ , 
 \label{eq:time}
 \end{equation} 
 with $N_c \sim 10^{48}$ being the number of nucleation centers expected in the innermost part 
 ($r \leq R_{nuc} \sim100$ m) of the hadronic star, where pressure and temperature 
 can be considered constant and equal to their central values. 
 
 The thermal nucleation rate can be written \cite{LanTur73} as 
 \begin{equation}
     I =\frac{\kappa}{2 \pi} \Omega_0 \exp (- U({\cal R}_c, T) /T)
 \label{eq:therm_rate}
 \end{equation}
 where the statistical prefactor (see Ref.\ \cite{CseKap92}), is given by: 
 \begin{equation}
    \Omega_0 = 
 \frac{2}{3\sqrt{3}} \Big(\frac{\sigma}{T}\Big)^{3/2} \Big(\frac{\cal R}{\xi_Q}\Big)^4 \ .
 \label{eq:omega}
 \end{equation}
  $\xi_Q$ is the quark correlation length, which gives a measure of the thickness of the 
 interface layer between the two phases (the droplet "surface thickness"). 
 In the present calculation we take $\xi_Q = 0.7$~fm according to the estimate given in 
 Refs.\ \cite{CseKap92,hei95}. 
 
 For the dynamical prefactor we have used a general expression which has been derived by 
 Venugopalan and Vischer \cite{VenVis94} (see also Refs.\ \cite{CseKap92,CKO03}) 
 \begin{equation}
 \kappa = \frac{2 \sigma} {{\cal R}_c^3 (\Delta w)^2} \Big [ \lambda T + 2 \Big(\frac{4}{3} \eta + \zeta \Big)\Big] \, ,
 \label{eq:kappa}
 \end{equation}
 where $\Delta w = w_{Q*} - w_H$ is the difference between the enthalpy density of the two phases, 
 $\lambda$ is the thermal conductivity, and $\eta$ and $\zeta$ are, respectively, the shear and bulk viscosities
 of hadronic matter. 
 According to the results of Ref.\ \cite{Dan84}, the dominant contribution to the prefactor 
 $\kappa$ comes from the shear viscosity $\eta$. Therefore, we take $\lambda$ and $\zeta$ equal to zero,
 and we use for the shear viscosity the following relation \cite{Dan84}: 
 \begin{equation}
 \eta = \frac{7.6 \times 10^{26}} {(T/{\rm MeV})^2} \Big(\frac{n_H}{n_0}\Big)^2 ~~ 
             \frac{{\rm MeV}}{{\rm fm \, s}} \, ,
 \label{eq:eta} 
 \end{equation} 
 with $n_0 = 0.16$~fm$^{-3}$ being the saturation density of normal nuclear matter. 
  The thermal nucleation time $\tau_{th}$, relative to the innermost stellar region 
 ($V_{nuc} = (4 \pi/3) R_{nuc}^3$) where almost constant pressure and temperature occur, can thus 
 be written as $\tau_{th} = (V_{nuc} \, I )^{-1}$.

  
\section{Results and discussion}
\label{sec:results}
 
\begin{figure}
 \centering
 \vspace{2cm}
 \begin{tabular}{c}
 \includegraphics[width=0.6\linewidth]{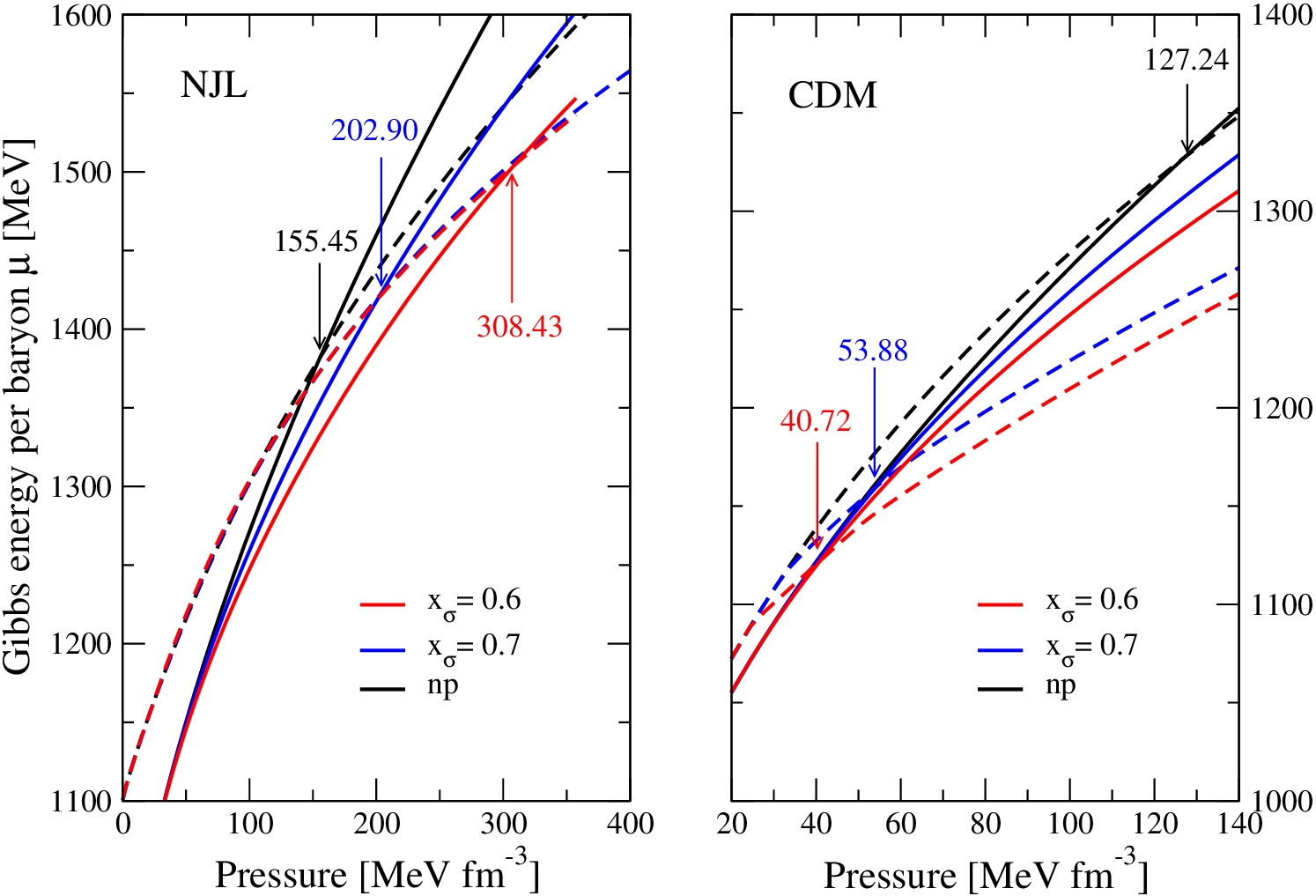}\\
 \end{tabular}
 \caption{(Color online) Gibbs energy per baryon at zero temperature ($T = 0$) 
 as function of the pressure for the hadronic (continuous lines) and the $Q^{*}$ (dashed lines) phase. 
 The arrows and the 
 corresponding numbers indicate the value of the pressure $P_0$ at which the bulk phase transition 
 takes place. Results for the NJL (CDM) model are shown in the left (right) panel.
 See text for more details. }
 \label{fig1}
 \end{figure}

\begin{figure}
 \centering
 \vspace{2cm}
 \begin{tabular}{c}
 \includegraphics[width=0.6\linewidth]{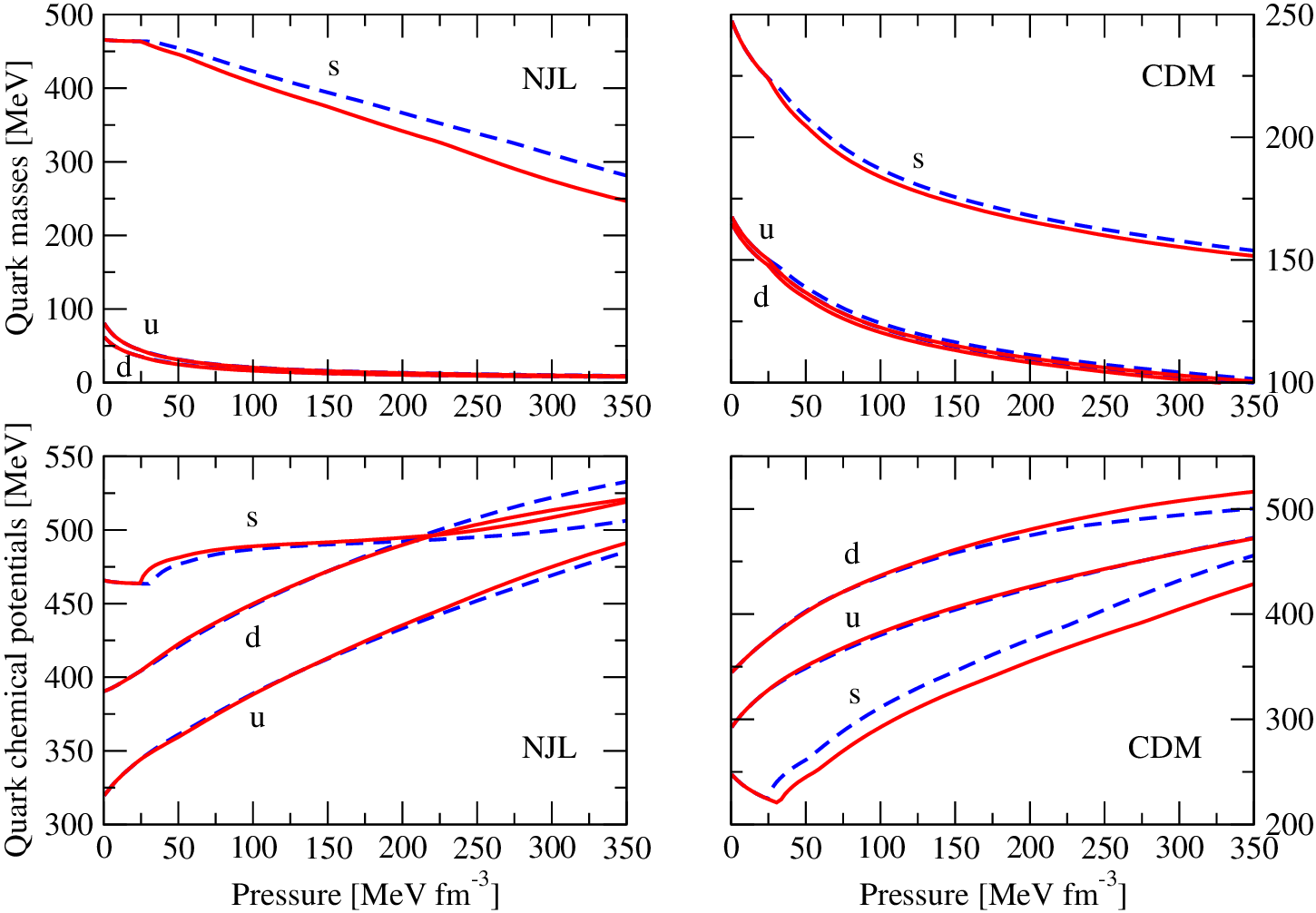}\\
 \end{tabular}
 \caption{(Color online) Quark masses (upper panels) and chemical potentials (lower panels) in the $Q^{*}$ phase at $T = 0$ 
using the NJL (left panels) and the CDM (right panels) models 
to describe the quark deconfined phase.   
Results for $x_{\sigma}=0.6$ and $x_{\sigma}=0.7$ are shown by the red continuous lines and the blue dashed ones, respectively.}
 \label{fig3}
 \end{figure}

In Fig.\ \ref{fig1} we show the Gibbs energy per baryon for the $\beta$-stable hadronic phase (continuous lines) and the $Q^{*}$-phase (dashed lines)
as a function of the pressure for various parametrizations of the hadronic equation of state
and the two models, NJL (left panel) and CDM (right panel), used to describe the deconfined phase. 
Above the phase equilibrium pressure $P_0$ the $\beta$-stable hadron phase is metastable, and the formation of the stable (with respect to the strong interactions) Q*-phase 
will occur via  a nucleation process.  
An interesting difference exists between the NJL and the Chromo Dielectric model: 
starting with a hadronic phase made of $\beta$-stable nucleonic matter (continuous line labeled np) 
and next including hyperons with an increasing concentration (continuous lines labeled  
respectively $x_\sigma = 0.7$, and  $x_\sigma = 0.6$), the transition point $P_{0}$ moves to 
lower values in the CDM, whereas in the NJL model  the opposite behaviour is observed.  
This ultimately can be traced back to the different numerical values and density (pressure) 
dependence of the dynamical strange quark mass in the two models. 
To elucidate this connection, we plot in Fig.\ \ref{fig3} the masses (upper panels) and chemical potentials (lower panels) of the $u,d$ and $s$ quarks in the non-$\beta$-stable $Q^*$-phase as function of the pressure for both models and for two different parametrizations of the hadronic phase ($x_\sigma=0.6,$ and 0.7).   
Here we note that the strange quark mass and chemical potential are 
much larger in the NJL model than in the CDM one in all range of the pressure explored,  
and particularly at the phase transition pressure $P_0$.  
 

It is important to note that a) at $P$=0 the density of
quarks $u$ and $d$ is not zero,  $\rho_d\sim 2\rho_u$. Therefore  the chemical potential is larger
than the effective mass and it is larger for the $d$ quarks;
b) for $P<$ 20-30 MeV fm$^{-3}$ the density of  $s$-quarks is zero and  the
s quark chemical potential coincides with the effective $s$ quark mass, the
chemical potential decreases with the pressure because the effective mass
decreases;  c) above $P>$ 20-30   MeV fm$^{-3}$ the $s$ quark density is nonzero and the chemical
potential increases with the pressure. The s-quark chemical potential is
smaller than the $u$ and $d$ quark chemical potential, although the mass is
larger, because the $u$ and $d$ quark densities are much larger. 
At 350 MeV fm$^{-3}$ the fractions of quarks $u$, $d$ and $s$ are respectively 0.33, 0.45 and 0.22. Comparing with the
NJL model, the big difference is the much larger effective mass the
$s$-quark has in NJL: the onset of the $s$-quark occurs at  the same pressure (set by the hyperon threshold in the $\beta$-stable hadronic phase) but the mass is much larger. 

In order to explain the different behaviour of the two chiral approaches, we observe that there are two opposite effects that define the phase transition: a smaller $x_\sigma$ gives rise to a larger strangeness fraction in the hadronic phase and, therefore, to a softer hadronic EOS, 
 shifting possible deconfinement transition point to larger densities in $\beta$-stable matter. 
 On the other hand, the $Q^*$ phase has the same strangeness content of the hadronic one: in the CDM model all quark masses are 
 similar and a more symmetric $uds$ quark matter is energetically favoured, while in the NJL model the $s$ quark mass is much larger and a larger $ud$ quark fraction is favoured. 
 This is confirmed comparing the quark chemical potential in both models. One can see that in the NJL model an increase of the hyperonic content of the hadronic phase gives rise just to a small decrease of the strange quark chemical potential, much smaller than the one occuring in the CDM model.  
 
\begin{figure}
 \centering
 \vspace{2cm}
 \includegraphics[width=0.6\linewidth]{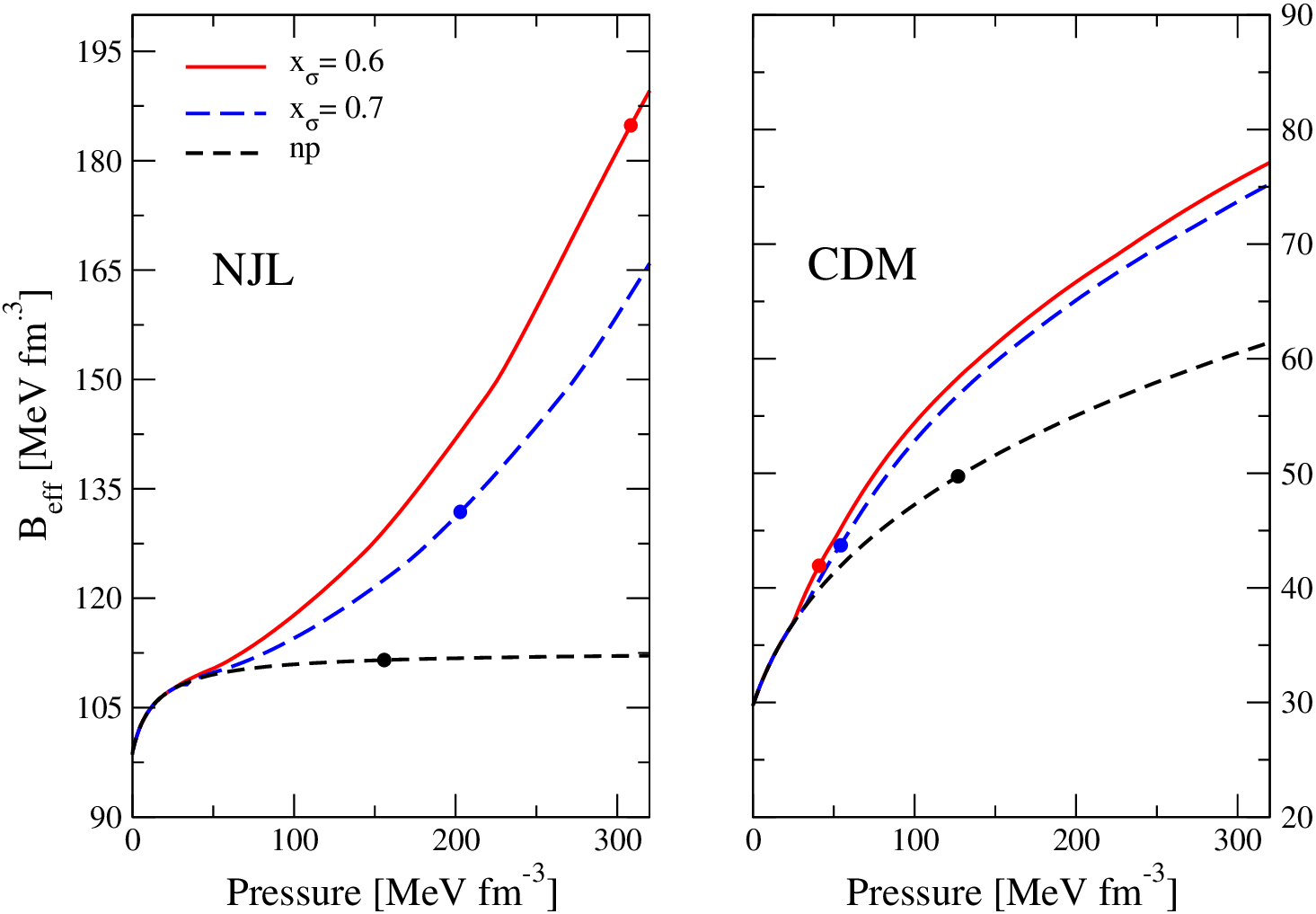}
 \caption{(Color online) Effective bag pressure at $T = 0$  
for the NJL (left panel) and the CDM (right panel) models.  
Results for $x_{\sigma}=0.6$, $x_{\sigma}=0.7$ and the pure nucleonic matter case ($np$) are shown. The filled circle denotes the value of the effective bag pressure at the 
transition point $P_0$.}
 \label{beff}
 \end{figure}

\begin{figure}
 \centering
 \vspace{2cm}
 \includegraphics[width=0.6\linewidth]{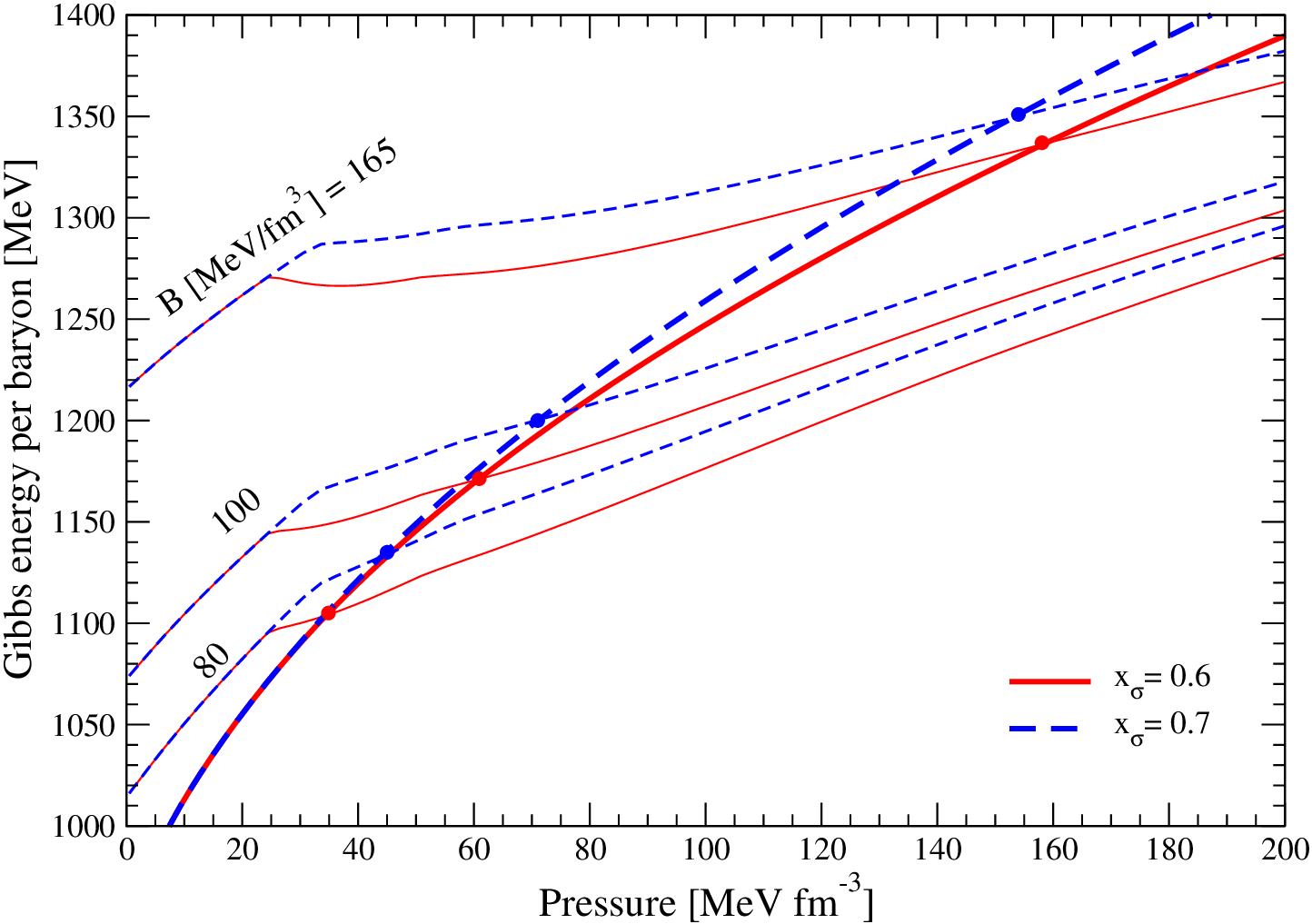}
 \caption{(Color online) Gibbs energy per baryon at $T = 0$ for the 
$\beta$-stable hadronic phase, with $x_\sigma=0.6$ (continuous thick line)
and $x_\sigma=0.7$ (dashed thick line),  and for the  $Q^{*}$ phase (thin lines).   
The  MIT bag model has been used for the $Q^{*}$ phase with a bag constant  
$B= 80,\, 100$ and 165 MeV/fm$^3$. } 
 \label{chem1}
 \end{figure}

It is also useful to analyse the effective bag pressure $B_{eff}$ for NJL and CDM models as a 
function of the pressure (Fig.\ \ref{beff}). The value of  $B_{eff}$ at the transition pressure $P_0$ is  denoted with a filled  circle.  
Coherently with the previous discussion, it is seen that for the NJL model the larger the hyperon content in the $\beta$-stable hadronic phase the larger is the value of $B_{eff}$ 
at the transition point. The opposite behaviour occurs for the CDM model. 
This is easly understood considering what happens in the MIT bag model for different values 
of the bag constant $B$. 
In this case $B_{eff}$ is a constant and the discussion carried out in this framework is more transparent.
 In Fig.\ \ref{chem1} the Gibbs energy per baryon for $Q^*$ (thin lines) and hadronic 
(thick lines) matter is plotted for $x_\sigma=0.6 $ and 0.7 and three values of the bag pressure $B$.  
Some conclusions are immediate: 
i) the larger the bag value the larger  the Gibbs energy per baryon for the 
Q* phase  therefore the transition is disfavoured, 
ii) the smaller the $x_\sigma$ value the smaller is the Gibbs energy per baryon both in the hadronic and in the $Q^*$ phase, 
iii) the last effect is much larger in  the $Q^*$ phase for the smaller pressures, but becomes of the same order of magnitude in both phases for $B >150$ MeV fm$^{-3}$. 
Taking together effect i) and iii)  and observing the very different value 
of $B_{eff}(P_0)$ in the two chiral approaches considered, we deduce that in the NJL model the phase  transition is disfavoured for the smaller value of $x_\sigma$ while the opposite occurs for the CDM model. 

\begin{figure}
 \centering
 \vspace{2cm}
 \begin{tabular}{c}
 \includegraphics[width=0.6\linewidth]{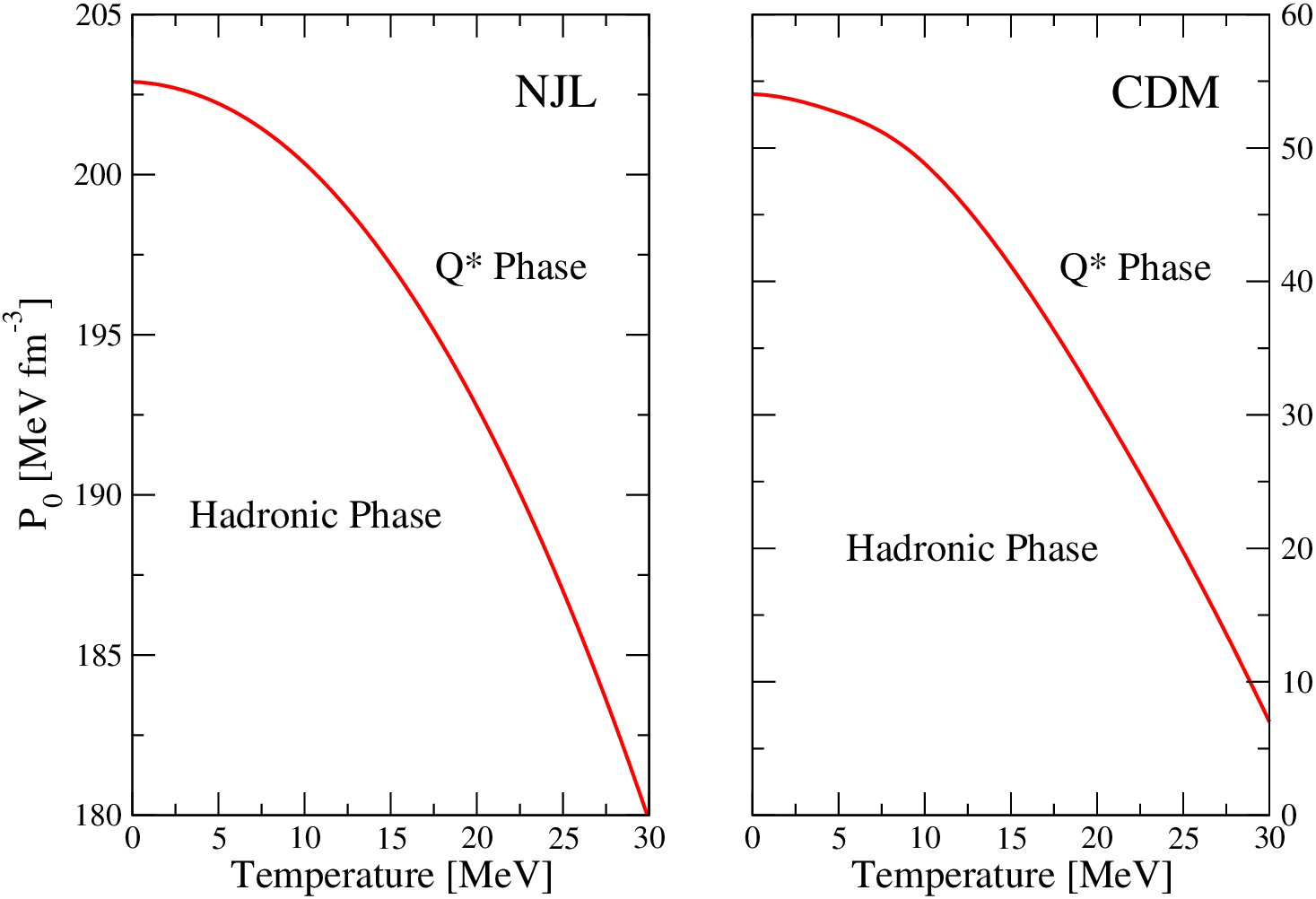}\\
 \end{tabular}
 \caption{(Color online) 
Phase equilibrium curve between the $\beta$-stable hadronic phase,  
with $x_{\sigma}=0.7$, and the $Q^{*}$ phase using the NJL (left panel) and the CDM (right panel) models.} 
 \label{fig6}
 \end{figure}

The phase equilibrium curve $P_{0}(T)$ between the hadron and  $Q^{*}$ phases 
is shown in Fig.\ \ref{fig6} in the case of the NJL model (left panel) or the CDM (right panel) 
to describe the deconfined phase. For the hadron phase we take $x_\sigma = 0.7$ in both cases.    
The region of the  $P_0$--$T$ plane above each curve represents the deconfined Q*-phase.  
We have checked that in the CDM, the phase separation line moves downwards 
(upwards) if an hadronic EOS with a larger (smaller) strangeness content is used, {\it e.g.,} 
$x_\sigma=0.6$ ($np$ matter). Using the NJL model the opposite occurs due to the reasons 
discussed above for the case of zero temperature phase transition.

 Now we can proceed to analyse the effects of quark matter nucleation process in the core of 
neutron stars.  We first consider quark matter nucleation in 
cold $\beta$-stable hadronic matter that is the typical situation in  the core of a neutron star a few minutes after its birth. 
At this point of our discussion, we assume that quark matter nucleation has 
not occured during the proto-neutron star stage (see below).      

In Fig.\ \ref{fig4}  we show the  gravitational mass versus central pressure for different  
compact star models. Hadronic star sequences are calculated using  the GM1 parametrization 
for pure nucleonic matter (black curve),  hyperonic matter with $x_{\sigma}=0.7$ (blue curve) 
and $x_{\sigma}=0.6$ (red curve). The quark star (QS) sequence is represented by the green curve.
For the quark models cosidered in this paper, all QS sequences are made of hybrid stars (YS).  
Results in the left (right) panel are relative to the NJL (CDM) model for the quark phase.  
The configuration marked with an asterisk represents, in all cases, the hadronic star for 
which the central pressure is equal to $P_0$ and thus the quark matter nucleation time 
is $\tau = \infty$.  The critical mass configuration is denoted by a full circle.  
The stellar conversion process \cite{b0,b1,b2} of the critical mass configuration into a 
final quark star with the same stellar baryonic mass (filled square) is denoted by the dashed 
line connecting the the circle to the square.  
Notice that in most of the cases reported in the figure the quark matter nucleation process 
will lead to the formation of a black hole (for these cases we do not plot in Fig.\ \ref{fig4} 
the coresponding YS sequence). In particular, within the present values for the EOS parameters 
the formation of quark stars is not possible modeling the quark phase with the NJL model.  

We next consider the case of new born hadronic stars (proto-hadronic stars, PHSs). 
In this case the quark deconfinement phase transition is likely triggered by a thermal 
nucleation process and it will occur in those PHSs with a gravitational mass 
$M > M_{cr}(\tilde{S})$ \cite{me,me1}.   
Here we consider the case of neutrino free matter, since it has been shown \cite{me1}
that neutrino trapping does not alter substantially the outcomes of the PHS evolution.

The evolution of a PHS within this scenario is delineated in Fig.\ \ref{fig9},  
where we plot the appropriate stellar equilibrium sequences 
in the gravitational--baryonic mass plane obtained from the CDM  for the quark phase and 
the GM1 model in the case of  hyperonic matter 
with $x_{\sigma}=0.6$ (left panel), and $x_{\sigma}=0.7$ (right panel).   
In particular, we plot the PHS sequence, {\it i.e.} isoentropic hadronic stars   
($\tilde {S} = 2~k_B$) and neutrino-free matter (upper line), 
and the cold hadronic star (HS) sequence (dashed line).  
The asterisk and the full circle on these lines identify respectively the stellar configuration 
with $\tau = \infty$ and the critical mass configuration.   
We denote as $M_{B,cr}^{PHS} \equiv M_{B,cr}(\tilde S = 2 k_B)$ the baryonic critical mass 
for the PHS sequence and as $M_{B,cr}^{HS} \equiv M_{B,cr}(\tilde S = 0)$ the 
baryonic critical mass for the cold hadronic star sequence.   
The lower continuous (green) line represents the cold QS sequence having a 
a maximum gravitational (baryonic) mass $M_{max}^{QS}$ ($M_{B,max}^{QS}$).   
We assume \cite {grb} ( $M_B =$~constant during these stages of the stellar evolution ).
We note, indeed, that sizeable mass accretion on the proto-neutron star occurs within a time of 
$\sim 0.5$~s after core bounce \cite{BurLat86,prak97}). During the subsequent stages,  
the star thus evolves with $M_B \simeq$~const. .  

Thus according to the results in the left panel of Fig.\ \ref{fig9},   
proto-hadronic stars with a baryonic mass  $M_B < M_{B,cr}^{PHS}$ 
($=1.16~M_\odot$ within the selected EOS parametrization)  
will survive Q*-matter {\it early nucleation}  ({\it i.e.} nucleation within the cooling time  
$t_{cool} \sim$~a~few~$10^2$~s) and in the end will form stable ($\tau = \infty$) cold hadronic stars.    
Proto-hadronic stars with $M_{B,cr}^{PHS} < M_B < M_{B,max}^{QS}$ ($= 1.79~M_\odot$ 
for the present EOS) will experience early nucleation of a Q*-matter drop and will ultimately 
form a cold deleptonized quark star.   
The last possibility is for PHSs having  $M_B >  M_{B,max}^{QS}$. In this case the early nucleation 
of a Q*-matter drop will trigger a stellar conversion process  to a cold QS configuation with 
$M_B > M_{B,max}^{QS}$, thus these PHSs will finally form  black holes. 
A similar evolutionary path is found in the case of $x_{\sigma}=0.7$ (right panel).

In Fig.\ \ref{fig8} we plot the PHS, cold HS, and cold YS sequences in the 
gravitational--baryonic mass plane for the case of the NJL model for the quark phase and 
the GM1 model in the case of pure nucleonic matter (right panel) or hyperonic matter 
with $x_{\sigma}=0.7$ (left panel).  It is clearly seen that in the case of the NJL model 
it is almost impossible to populate the YS branch. Cold quark stars can be formed  
in the case of $x_{\sigma}=0.7$ (left panel) for a very narrow range of baryonic stellar 
masses $ 2.20 <  M_B/M_\odot < 2.23 M_\odot$.

Finally, in tables \ref{t1}  and \ref{t2}  we report the values of the gravitational and baryonic 
critical mass for the PHS and HS sequences for the two adoped quark matter models.     
As it has been found in previous works \cite{me,me1}, thermal effects reduce the values 
of the critical mass and increase the portion of the quark star branch that can populated via 
the stellar conversion process \cite{grb,b0,b1}.  
Notice that the maximum mass configuration of the YS sequence is insensitive, in the case of the 
CDM, to the value of the hyperon coupling $x_\sigma$, since in this case the threshold density for 
quark deconfinement is much lower than the density for the onset of hyperons.

 \section{Conclusion}
 \label{sec:conclusions}

In this paper we have studied the nucleation of quark matter in both hot and cold $\beta$-stable hadronic matter using two different models with chiral symmetry to describe the quark phase: 
the Nambu--Jona-Lasinio model and the Chromo Dielectric model. 
For the hadronic phase we chose the GM1 parametrization of the non-linear Walecka model and 
we have considered pure nucleonic matter as well as hyperonic matter with a large hyperon  
fraction ($x_\sigma$=0.6), and a small hyperon fraction ($x_\sigma=0.7$).

 The nucleation process forms a short-lived transitory phase ($Q^*$ phase) which has the same 
flavor content of the initial $\beta$-stable hadronic phase. 
This particular circumstance, together with the different pressure (density) dependence of 
the strange quark effective mass in the two employed quark matter models  
produces considerable differences on the bulk properties of the phase transition and on 
neutron star composition and early evolution.  
More precisely, we found that for the NJL model the presence of hyperons disfavour the phase 
transition pushing  the transition point to very high densities while with the CDM the opposite 
behaviour has been observed.  In addition, we found that in the case of the NJL model it is almost 
impossible to populate the quark star branch and that quark matter nucleation will lead 
to the formation of a black hole. Thus within the NJL model for the quark phase, all compact stars 
are pure hadronic stars.   
In the case of the CDM, thermal effects reduce the value of the critical mass,  
and both hadronic and quark star configurations can be formed as a result of the evolution 
of  proto-hadronic stars, depending on the value of the stellar baryonic mass.

 A very recent measurement \cite{Demorest10} of the mass of the pulsar PSR J1614-2230 makes 
it the most massive neutron star known to date with a mass  M = (1.97 $\pm$ 0.04) $M_\odot$. 
Within the EOS models employed in the  present work, the compact star in  PSR J1614-2230 
could only be a pure HS (in the case the quark phase is described by the  NJL model, Fig.\ \ref{fig8})  
formed from the evolution of a PHS with initial baryonic mass $M_B < M_{B,cr}^{PHS}$ and  
after a long-term mass accretion ($M_{accr} \sim $ 0.1 -- 0.2 $M_\odot$) from a companion star 
in a binary system. This long term evolution can finally form pure hadronic star with a mass  
$M < M_{cr}^{HS}$, with  $M_{cr}^{HS}   = 2.025 M_\odot$ (case $x_\sigma = 0.7$) or   
$M_{cr}^{HS}   = 2.238 M_\odot$ (case of np matter).     
The CDM model fails to predict  such a high mass.

  
\section*{Acknowledgments}
  This work has been partially supported by FEDER/FCT (Portugal) 
 under grants SFRH/BD/62353/2009, PTDC/FIS/113292/2009 and
 CERN/FP/116366/2010,
 by the Ministero dell'Universit\`a e della Ricerca (Italy) under the 
 PRIN 2009 project {\it Theory of Nuclear Structure and Nuclear Matter},
 and by COMPSTAR, an ESF Research Networking Programme.


 \newpage
 \begin{table} 
 \begin{center}
 \bigskip                           
 \begin{ruledtabular}
 \begin{tabular}{l|ccccccc}
           & $M_{cr}^{HS}$  & $M_{B,cr}^{HS}$ & $M_{cr}^{PHS}$ & $M_{B,cr}^{PHS}$ & $M_{max}^{YS}$ &  $M_{B, max}^{YS}$  \\
 \hline                     
  $x_{\sigma}$ = 0.7  &  2.025  &  2.342  &  1.964  &  2.201  & 1.943  &  2.293  \\    
\hline                       
  $np$                      & 2.238  &  2.634  &   2.112 &  2.386  &  1.988  &  2.287  \\ 
                       
 \end{tabular}
 \end{ruledtabular}
 \end{center} 
 \caption{
Gravitational $M_{cr}^{HS}$ ($M_{cr}^{PHS}$) and baryonic  $M_{B,cr}^{HS}$ 
($M_{B,cr}^{PHS})$  critical mass for the cold hadronic star (proto-hadronic star) sequence.  
The gravitational and baryonic maximum mass for the cold hybrid star sequence are denoted respectively as  
$M_{max}^{YS}$ and $M_{B, max}^{YS}$.  
The values of stellar masses  are in unit of the solar mass ($M_\odot=1.989 \times 10^{33}$ g). 
All the results reported in the table are relative to the GM1 equation of state for the 
hadronic phase and the Nambu--Jona-Lasinio (NJL) model for the quark phase. 
The gravitational maximum mass for the cold hadronic star sequence is   
$M_{max}^{HS} = 2.042 M_\odot$ in the case of hyperonic matter with $x_\sigma = 0.7$, and 
$M_{max}^{HS} = 2.364 M_\odot$ in the case of nucleonic (np) matter.
}
 \label{t1}
 \end{table}
 
\newpage 
 \begin{table} 
 \begin{center}
 \bigskip
 \begin{ruledtabular}
 \begin{tabular}{l|ccccccc}
  $x_{\sigma}$  & $M_{cr}^{HS}$  & $M_{B,cr}^{HS}$ & $M_{cr}^{PHS}$ & $M_{B,cr}^{PHS}$ & $M_{max}^{YS}$ & $M_{B, max}^{YS}$  \\
 \hline   
      0.6  & 1.453 & 1.604 &  1.092 & 1.153  &  1.557  &  1.793 \\                     
 \hline
      0.7 & 1.741 & 1.963 & 1.204 & 1.279    &  1.557  &  1.793 \\   
                      
 \end{tabular}
 \end{ruledtabular}
 \end{center} 
 \caption{
Same as in table \ref{t1} but using the CDM for 
the quark phase.
The gravitational maximum mass for the cold hadronic star sequence is 
$M_{max}^{HS} = 1.791 M_\odot$ in the case $x_\sigma = 0.6$, and  
$M_{max}^{HS} = 2.042 M_\odot$ in the case $x_\sigma = 0.7$. 
}
\label{t2}
\end{table}
 

 \newpage

\newpage

 \begin{figure}
 \centering
 \vspace{2cm}
 \includegraphics[width=0.5\linewidth]{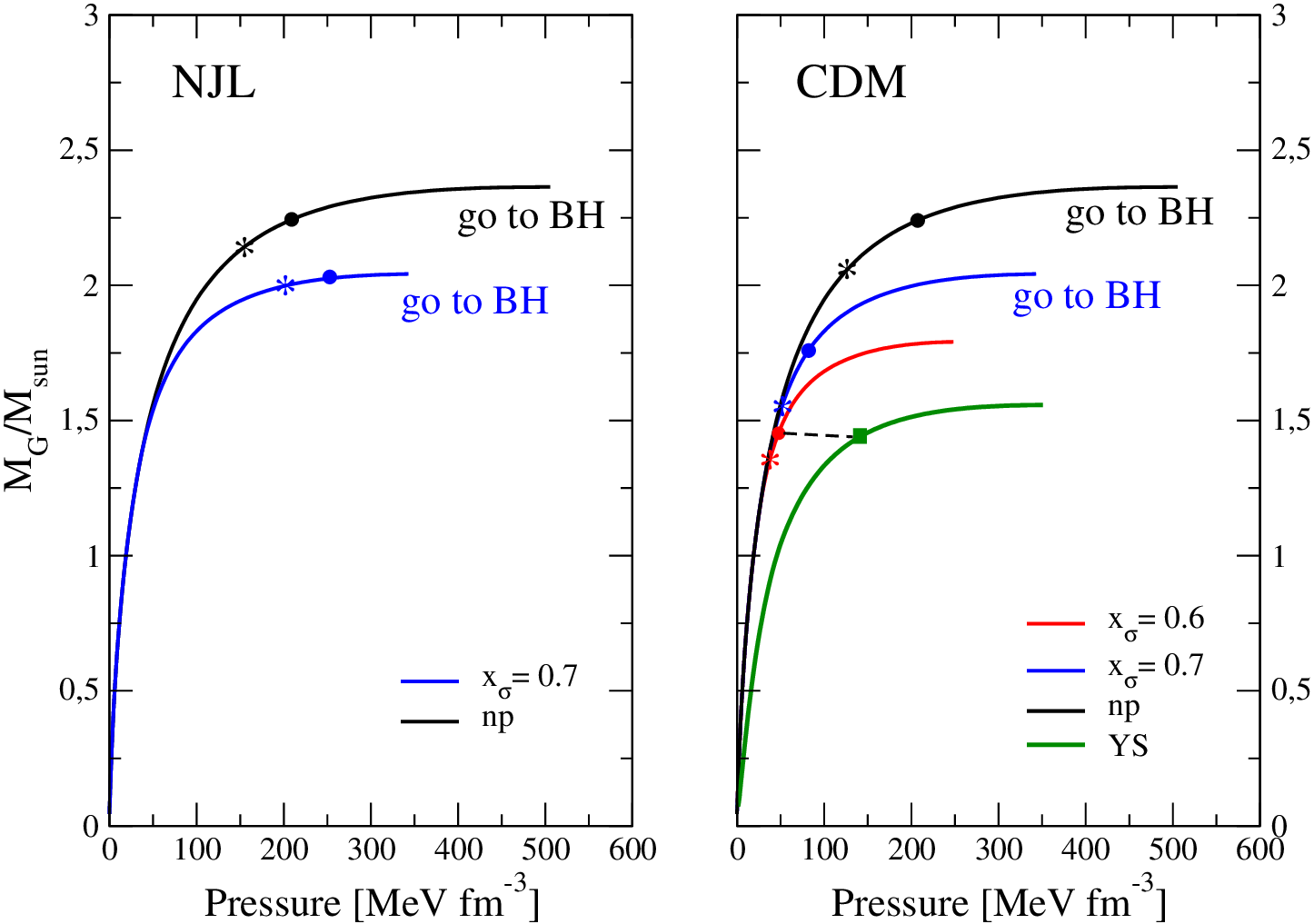}
 \caption{(Color online) 
Gravitational mass versus central pressure for compact stars.      
Hadronic star sequences are calculated using  the GM1 parametrization 
for pure nucleonic matter (black curve), hyperonic matter with $x_{\sigma}=0.7$ (blue curve) 
and $x_{\sigma}=0.6$ (red curve).   
The hybrid star (YS) sequence is represented by the green curve.  
Results in the left (right) panel are relative to the NJL (CDM) model for the quark phase.  
The configuration marked with an asterisk represents in all cases the hadronic star for 
which the central pressure is equal to $P_0$ and thus the quark matter nucleation time 
is $\tau = \infty$.  The critical mass configuration is denoted by a full circle.   
The conversion process of the critical mass configuration into a final quark star with 
the same stellar baryonic mass (filled square) is denoted by the dashed line connecting the 
the circle to the square.  
In most of the cases reported in the figure the quark matter nucleation process will lead to the formation of a black hole (go to BH). For these cases we do not plot the coresponding YS sequence.
}
 \label{fig4}
 \end{figure}

 \begin{figure}
 \centering
 \vspace{2cm}
 \begin{tabular}{c}
 \includegraphics[width=0.6\linewidth]{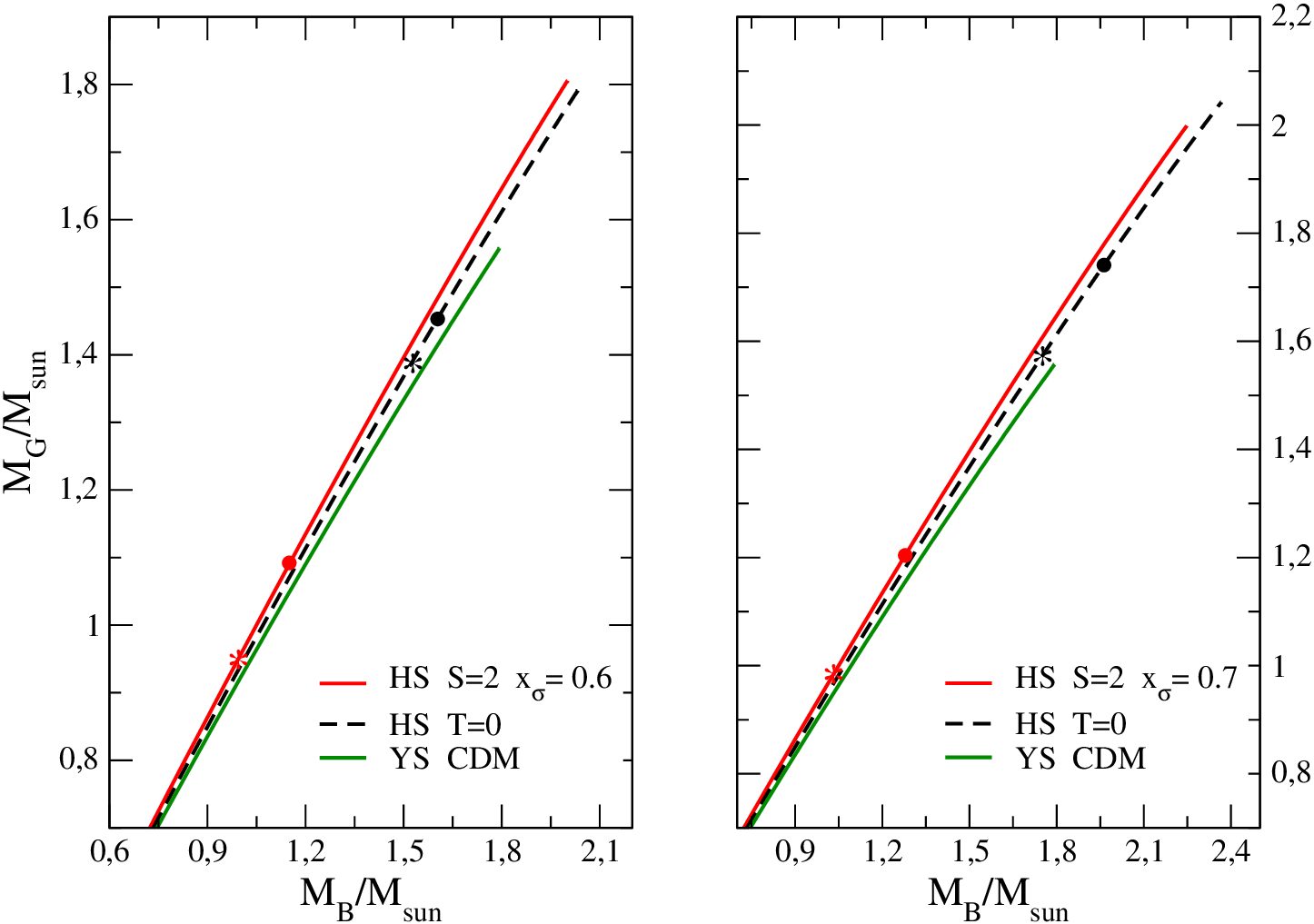}\\
 \end{tabular}
 \caption{
Evolution of a proto-hadronic star in the gravitational--baryonic mass plane 
using the Chromo Dielectric model (CDM) to describe the quark phase and the GM1 model 
with $x_{\sigma}=0.6$ (left panel) and $x_\sigma=0.7$ (right panel) for the hadronic phase.  
The upper (red) line represents the stellar equilibrium sequence for neutrino-free  
proto-hadronic stars (PHS) with $\tilde {S} = 2~k_B$.  
The (black) dashed line represents the cold the HS sequence. 
The asterisk and the full circle on these lines represent the stellar configuration with nucleation 
time $\tau = \infty$  and the critical mass configuration, respectively.  
The lower (green) line represent the cold YS sequence.    
Assuming  $M_B =~$const, the evolution of a PHS in this plane occurs along a vertical line.} 
 \label{fig9}
 \end{figure} 
 \newpage

 \begin{figure}
 \centering
 \vspace{2cm}
 \begin{tabular}{c}
 \includegraphics[width=0.6\linewidth]{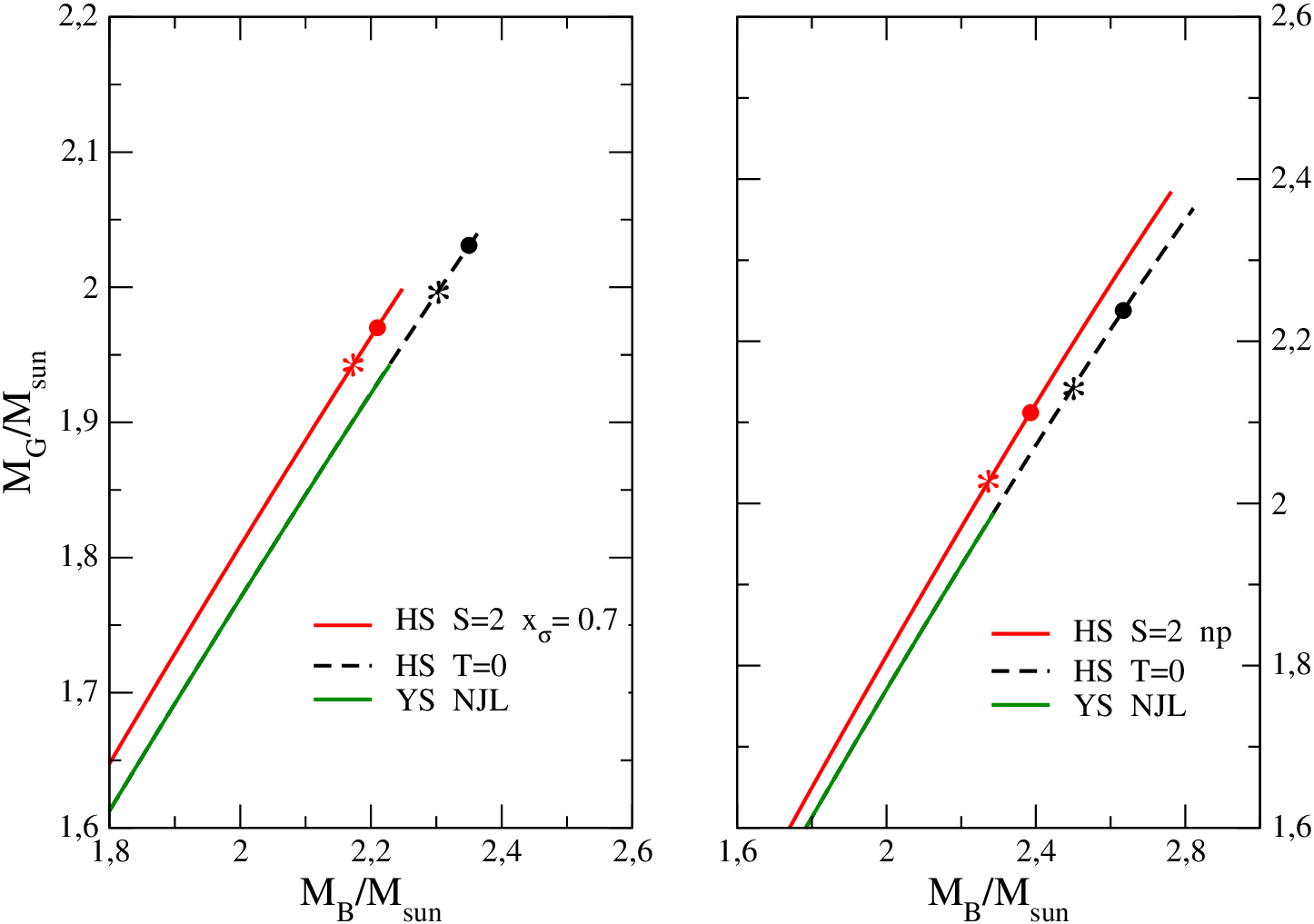}\\
 \end{tabular}

\caption{
Same as in the previous figure but in the case of the NJL model 
for the quark phase and the GM1 model for hyperonic matter with 
$x_{\sigma}=0.7$ (left panel) and pure nucleonic (np) matter (right panel).}   
 \label{fig8}
 \end{figure} 


\begin{thebibliography}{200}
 
\bibitem{b0} Z. Berezhiani, I. Bombaci, A. Drago, F. Frontera, A. Lavagno, 
                     {Nucl. Phys. B - Proceedings Supplements} {\bf 113} (2002) 269.
 
\bibitem{b1} Z. Berezhiani, I. Bombaci, A. Drago, F. Frontera, A. Lavagno, 
                            {Astrophys. Jour.} {\bf 586} (2003) 1250.
 
\bibitem{b2} I. Bombaci, I. Parenti, I. Vida\~na, {Astrophys. Jour.} {\bf 614}, (2004) 314.
 
\bibitem{b3} A. Drago, A. Lavagno, G. Pagliara, Phys. Rev. {\bf D69} (2004) 057505. 
 
\bibitem{b4} G. Lugones, I. Bombaci, Phys. Rev. {\bf D72} (2005) 065021.
 
\bibitem{b5} I. Bombaci, G. Lugones, I. Vida\~na, {Astron. and Astrophys.} {\bf 462} (2007) 1017.
 
\bibitem{b6} I. Bombaci, P.K. Panda, C. Provid\^encia, I. Vida\~na, Phys. Rev. {\bf D77} (2008) 083002.
 
\bibitem{b7} C. Bambi, A. Drago, {Astropart. Phys.} {\bf 29} (2008) 223.
 
\bibitem{b8} A. Drago, G. Pagliara, J. Schaffner-Bielich, J. Phys. {\bf G35} (2008) 014052. 
  
\bibitem{h1} J. E. Horvath, O. G. Benvenuto, H. Vucetich, {Phys. Rev.} {\bf D45} (1992) 3865.
 
\bibitem{h2} J. E. Horvath, {Phys. Rev. } {\bf D49} (1994) 5590.
 
\bibitem{h3} M. L. Olesen and J. Madsen, Phys. Rev. {\bf D49} (1994) 2698.
 
\bibitem{h4} H. Heiselberg, in Strangeness and Quark Matter, Ed. G. Vassiliadis, World Scientific (1995) 338; arXiv:hep-ph/9501374. 

\bibitem{me}  I. Bombaci, D. Logoteta, P.K. Panda, C. Providencia and I. Vida\~na Phys. Lett. {\bf B680} 448 (2009). 
 
\bibitem{me1} I. Bombaci, D. Logoteta, C. Providencia and I. Vida\~na Astron. and Astrophys. {\bf 528}, A71 (2011). 
 
\bibitem{mit}   E. Farhi, R. L. Jaffe, Phys. Rev. {\bf D30}  2379 (1984).

\bibitem{lug1} G. Lugones, T.A.S. do Carmo, A.G. Grunfeld, N.N. Scoccola Phys. Rev. {\bf D81} 085012 (2010).
 
\bibitem{nambu} Y. Nambu and G. Jona-Lasinio, Phys Rev. {\bf 122} 345 (1961).
 
\bibitem{cdm} H. J. Pirner, G. Chanfray and O. Nachtmann, Phys. Lett. {\bf B147} 249 (1984).

\bibitem{bba2} M. Buballa, Pyhs. Rep. {\bf 407} 205 (2005)

\bibitem{bbub} M. Buballa, F. Neumannb, M. Oertel and I. Shovkovyd Phys. Lett. {\bf B595} 36 (2004).
 
\bibitem{cost} D.P. Menezes and C. Provid\^encia, {\em Phys. Rev.} {\bf C68} 035804 (2003).
 
\bibitem{bba} M. Baldo, G. F. Burgio, P. Castorina, S. Plumari and D. Zappala' Phys. Rev. {\bf C75} 035804 (2007)
 
\bibitem{bba1} M. Baldo, M. Buballa, G.F. Burgio, F. Neumann, M. Oertel and H.-J. Schulze Phys. Lett. {\bf B562} 153 (2003)
 
\bibitem{benh1} O. Benhar and A. Cipollone, Astron. and Astrophys.  {\bf 525} L1 (2011)

\bibitem{bla} D. Blaschke, S. Fredriksson, H. Grigorian, A.M. Oztas and F. Sandin Phys. Rev. {\bf D72} 065020 (2005).
 
\bibitem{rust} S. B. Ruester, V. Werth, M. Buballa, I. A. Shovkovy and D. H. Rischke Phys. Rev. {\bf D72} 034004 (2005).

\bibitem{njll} P. Rehberg, S.P. Klevansky and J. Hufner, Phys. Rev. {\bf C53} 410 (1996). 

\bibitem{lug} G. Lugones, A.G. Grunfeld, N.N. Scoccola, C. Villavicencio Phys. Rev. {\bf D80} 045017 (2009).

 
 \bibitem{gm91} N. K. Glendenning and S. Moszkowski, {\em Phys. Rev. Lett.} {\bf 67} 2414 (1991).
 
 \bibitem{prak97} M. Prakash, I. Bombaci, M. Prakash, P. J. Ellis, J. M. Lattimer and R. Knorren, {\em Phys. Rep.}{\bf 280} 1 (1997).
 
\bibitem{BurLat86}  
 A. Burrow, J. M. Lattimer,  {Astrophys. Jour.} {\bf 307} (1986) 178. 

 \bibitem{glen00} N. K. Glendenning, {\em Compact Stars}, (Springer-Verlag, New-York, 2000).
  
 \bibitem{bub} M. Buballa, F. Neumann, M. Oertel and I. Shovkovy Phys. Lett. {\bf B595} 36 (2004).
 
 \bibitem{cdm1} J. A. Mcgovern, M. C. Birse and D. Spanos, J. Phys. {\bf G16} 1561 (1990).
 
 \bibitem{cdm2} W. Broniowski, M. Cibej and M. Kutschera and M. Rosina Phys. Rev. {\bf D41} 285 (1990).
 
 \bibitem{cdm3} S. K. Gosh, and S. C. Pathak, J. Phys {\bf G18} 755 (1992).
 \bibitem{cdmf} T.Neuber, M.Fiolhais, K.Goeke and J.N.Urbano, Nucl.Phys. A 560, 909 (1993).
 \bibitem{cdm4} V. Barone and A. Drago, J. Phys. {\bf G21} 1317 (1995).
 
 \bibitem{cdmbaldo} C. Maieron, M. Baldo, G. F. Burgio and H.-J. Shulze Phys. Rev. {\bf D70} 043010 (2004).
 
 \bibitem{cdm5} W. M. Alberico, A. Drago and C. Ratti, N. Phys. {\bf A706} 143 (2002).
 
 \bibitem{cdm6} S. K. Gosh, S. C. Pathak and P. K. Sahu, Z. Phys {\bf A352} 457 (1995).
  
 \bibitem{cdm7} A. Drago, U. Tambini M. Hjorth-Jensen, Phys. Lett. {\bf B380} 13 (1996).
 
 \bibitem{cdm8} A. Drago, A. Lavagno, Phys. Lett. {\bf B511} 229 (2001).
 
 \bibitem{cdm9} A. Drago and E. Tambini, J. Phys. {\bf G25} 971 (1999).

 
 \bibitem{cdmqm}A. Drago, A. Fiolhais and U. Tambini. Nucl. Phys. A 588, 801 (1995)
 
 \bibitem{nlwm} B. D. Serot and J. D. Walecka, Adv. Nucl. Phys. 16, 1 (1986); J. Boguta and A. R. Bodmer, Nucl. Phys. A 292, 413 (1977).
 
\bibitem{fk04} 
             Z. Fodor, S. D.  Katz, {Prog. Theor.  Suppl.} {\bf 153} (2004) 86. 

\bibitem{hs98} S. D. H.  Hsu,  M. Schwetz, {Phys. Lett. B} {\bf 432} 
                                    (1998) 2003. 

 \bibitem{oli87} A. V. Olinto, Phys. Lett. {\bf B192} 71 (1987).
  
  \bibitem{hbp91} H. Heiselberg, G. Baym, C. J. Pethick, Nucl. Phys. B (Proc. Suppl.) 24 144 (1991).
  
  \bibitem{grb}  I. Bombaci and B. Datta, {\em Astrophys. J.} {\bf 530} (2000) L69.
 
  \bibitem{iida98} K. Iida and K. Sato, Phys. Rev. {\bf C58} 2538 (1998).
 
 \bibitem{hei93} H. Heiselberg, C.J. Pethick, E.F. Staubo, {Phys. Rev. Lett.} {\bf 70} 1355 (1993). 
 
 \bibitem{iida97} K. Iida, K. Sato, {Prog. Theor. Phys.} {\bf 98} 277 (1997). 
 
 \bibitem{lk72} I. M. Lifshitz, Y. Kagan, {Sov. Phys. JETP} {\bf 35} 206 (1972).
 
 \bibitem{LanTur73} J.S. Langer, L.A. Turski, Phys. Rev. {\bf A8} 3230 (1973).
 
 \bibitem{CseKap92} L. Csernai, J. I. Kapusta, {Phys. Rev.} {\bf D46} 1379 (1992). 
 
 \bibitem{hei95} H. Heiselberg, in: G. Vassiliadis (Ed.), Strangeness and Quark matter, World Scientific, 1995, p. 338 
 
 \bibitem{VenVis94} R. Venugopalan, A. P. Vischer, {Phys. Rev. } {\bf E49} 5849 (1994).
 
 \bibitem{CKO03} L. Csernai, J. I. Kapusta, E. Osnes, {Phys. Rev. } {\bf D67} 045003 (2003).
 
 \bibitem{Dan84} P. Danielewicz, {Phys. Lett. } {\bf B146} 168 (1984).
 
 \bibitem{Demorest10} P. B. Demorest, T. Pennucci, S. M. Ransom, H. S. E. Roberts, J. W. T. Hessel, Nature 467, 1081 (2010).
 
 \end{thebibliography}
  \end{document}